\documentclass[twocolumn, amssymb,  amsmath, amsfonts, nobibnotes, a4paper, aps, prd, superscriptaddress, nofootinbib]{revtex4-2}

\usepackage[T3,T1]{fontenc}
\DeclareMathAlphabet{\mathpzc}{OT1}{pzc}{m}{it} 
\usepackage{mathrsfs} 
\usepackage{bm} 
\makeatletter
\@ifclassloaded{beamer}
  { 
    \typeout{UsePackages: Detected beamer}
    \usepackage{tgheros}
    
  }
  { 
    \typeout{UsePackages: Did not detect beamer}
    \ifx\asybeamer\undefined 
    \typeout{UsePackages: Detected article}
    \usepackage[varg]{txfonts} 
    \usepackage{tgtermes} 
    \else 
    \typeout{Fonts: Detected asy for beamer}
    \usepackage{tgheros}
    
    \fi
  }

\usepackage{siunitx}
\usepackage{relsize}
\usepackage{booktabs}

\usepackage{multirow, newfloat}
\usepackage{graphicx} %
\usepackage[table,x11names,svgnames,rgb]{xcolor}
\usepackage{xspace}

\usepackage{mathtools}
\DeclareSymbolFontAlphabet{\mathrm}{operators}

\makeatletter %
\newcommand{\ShowFont}{%
  \typeout{The main font is \f@encoding \space \f@family \space %
    \f@series \space \f@shape \space at \f@size pt.}%
  \typeout{The math font sizes are \tf@size pt (main), \sf@size pt %
    (script), and \ssf@size pt (scriptscript).}%
  \typeout{The linewidth is \the\linewidth}} %
\makeatother %

\usepackage[colorlinks]{hyperref}
\usepackage{orcidlink}

\newcommand{\dphi}{\delta\hat{\varphi}}
\newcommand{\dpi}{\delta\hat{p}}

\usepackage[first=0,last=9]{lcg}

\DeclareRobustCommand{\modelstyle}[1]{\texttt{#1}}

\newcommand{\phX}{\mbox{\modelstyle{IMRPhenomX}}\xspace}

\newcommand{\phXHM}{\mbox{\modelstyle{PhenomXHM}}\xspace}
\newcommand{\phXPHMST}{\mbox{\modelstyle{PhenomXPHM-SpinTaylor}}\xspace}

\newcommand{\phXPHM}{\mbox{\modelstyle{PhenomXPHM}}\xspace}
\newcommand{\phXPNR}{\mbox{\modelstyle{PhenomXPNR}}\xspace}
\newcommand{\SEOBvFive}{\mbox{\modelstyle{SEOBNRv5PHM}}\xspace}

\newcommand{\bilby}{\mbox{\textsc{Bilby}}\xspace}

\newcommand{\bilbytgr}{\mbox{\textsc{Bilby TGR}}\xspace}

\newcommand{\dynesty}{\mbox{\textsc{Dynesty}}\xspace}
\newcommand{\numpy}{\mbox{\textsc{NumPy}}\xspace}
\newcommand{\scipy}{\mbox{\textsc{SciPy}}\xspace}
\newcommand{\matplotlib}{\mbox{\textsc{Matplotlib}}\xspace}

\DeclareMathAlphabet{\mathcalstd}{OMS}{cmsy}{m}{n}
\DeclareMathAlphabet{\mathpzc}{OT1}{pzc}{m}{it}

\DeclareRobustCommand{\sxsid}[1]{\texttt{SXS:BBH:#1}}

\newcommand{\UCLouvain}{University of Louvain, Centre for Cosmology, Particle Physics and Phenomenology---CP3, \\ Chemin du Cyclotron 2, 1348 Louvain-la-Neuve, Belgium}
\newcommand{\ROB}{Royal Observatory of Belgium, Avenue Circulaire, 3, 1180 Uccle, Belgium}

\newcommand{\Nikhef}{Nikhef, Science Park 105, 1098 XG Amsterdam, The Netherlands}

\newcommand{\VUAmst}{Department of Physics and Astronomy, Vrije Universiteit Amsterdam, 1081 HV Amsterdam, The Netherlands}

\definecolor{dodgerblue}{HTML}{1E90FF}
\hypersetup{citecolor=dodgerblue,  urlcolor=dodgerblue}

\definecolor{RED}{HTML}{F5054F}
\definecolor{LIGHT_ORANGE}{HTML}{FDAA48}
\definecolor{DARK_ORANGE}{HTML}{FF5B00}
\definecolor{LIGHT_BLUE}{HTML}{448EE4}
\definecolor{DARK_BLUE}{HTML}{0343df}
\definecolor{LIGHT_GREEN}{HTML}{40C53C}
\definecolor{DARK_GREEN}{HTML}{02590F}
\definecolor{rowshade}{gray}{0.91}

\newcommand{\Hz}{\ensuremath{\mspace{2mu}\text{Hz}}\xspace}

\newcommand{\flow}{\ensuremath{f_{\mathrm{low}}}\xspace}
\newcommand{\fref}{\ensuremath{f_{\mathrm{ref}}}\xspace}
\newcommand{\fhigh}{\ensuremath{f_{\mathrm{high}}}\xspace}

\newcommand\abs[1]{\ensuremath{\lvert#1\rvert}}

\newcommand\inp[2]{\langle #1 \,|\, #2 \rangle}

\def \msun  {\rm{M}_\odot}

\begin{document}

\title{Highly spinning binary black holes: promise and pitfalls for parameterized tests of general relativity}

\author{Soumen Roy\orcidlink{0000-0003-2147-5411}}
\email{soumen.roy@uclouvain.be}
\affiliation{\UCLouvain}
\affiliation{\ROB}

\author{Maria Haney\orcidlink{0000-0001-7554-3665}}
\affiliation{\Nikhef}
\affiliation{\VUAmst}

\author{Justin Janquart\orcidlink{0000-0003-2888-7152}}
\affiliation{\UCLouvain}
\affiliation{\ROB}

\begin{abstract}

Highly spinning binary black holes exhibit complex inspiral dynamics and offer a powerful opportunity to probe gravity in a stronger-field and more dynamical regime than non-spinning binaries. Gravitational-wave signals from such systems enable stringent parameterized tests of deviations from the source dynamics predicted by general relativity. We show that highly spinning, mass-asymmetric binaries can provide enhanced sensitivity to deviations entering at higher post-Newtonian orders, which remain comparatively poorly constrained. 
%
Using NR simulations from the SXS catalog, we first recover the injected signals within general relativity using \texttt{IMRPhenomXPHM-SpinTaylor}, \texttt{IMRPhenomXPNR}, and \texttt{SEOBNRv5PHM}, finding model-dependent biases in the inferred masses, spins, and spin orientations. We then analyze the same signals with the \texttt{IMRPhenomXPHM}-based inspiral parameterized test of general relativity framework, \textsc{TIGER}, and find that these waveform-systematic biases can lead to apparent deviations from general relativity, which become significant for loud signals. Our results highlight that highly spinning binaries offer powerful tests of higher-order post-Newtonian deviations, but waveform inaccuracies can lead to false signs of new physics.


\end{abstract}
\maketitle

\section{Introduction}

The LIGO–Virgo–KAGRA (LVK) gravitational-wave (GW) detector network~\cite{LIGOScientific:2014pky, VIRGO:2014yos, KAGRA:2020tym} has reported over 390 events through the second part of the fourth observing run (O4b)~\cite{LIGOScientific:2026wfs, LIGOScientific:2025slb, KAGRA:2021vkt, LIGOScientific:2020ibl, LIGOScientific:2018mvr, Nitz:2021zwj, Wadekar:2023gea}, all of which are produced by mergers of compact binaries. The fourth observing run concluded in November 2025 with a
third part (O4c), whose candidates have so far been released as public
alerts~\cite{GraceDBO4}. A six-month intermediate run (IR1) is planned for early
2027~\cite{LVK_observing_plans}, before the fifth observing run is expected to deliver a significant boost in sensitivity~\cite{KAGRA:2013rdx}. These detections, comprising mergers of binary black hole (BBH), binary neutron star (BNS), and neutron star–black hole (NSBH) systems, have opened new avenues for studying compact objects and the environments in which they form and evolve.

Rapidly spinning BBHs can exhibit complex dynamics during the inspiral, including strong spin-orbit interactions and precession of the orbital plane, which imprint characteristic amplitude and phase modulations on the observed signal~\cite{Kidder:1992fr, Apostolatos:1994mx, Kidder:2007rt}. These features, together with the merger and ringdown dynamics, provide a powerful opportunity to probe gravity in the strong-field, highly dynamical regime~\cite{LIGOScientific:2016lio, Yunes:2013dva}. At the same time, accurately modelling such systems remains challenging~\cite{MacUilliam:2024oif}, particularly when the spins are large and misaligned with the orbital angular momentum.

GW observations of compact-binary coalescences enable a variety of tests of general relativity (GR)~\cite{LIGOScientific:2016lio, LIGOScientific:2018dkp, LIGOScientific:2019fpa, LIGOScientific:2020tif, LIGOScientific:2021sio, LIGOScientific:2026qni, LIGOScientific:2026fcf, LIGOScientific:2026wpt, LIGOScientific:2026oim}. Among them, parameterized tests modify selected coefficients of the waveform phase and infer these modifications directly from the data~\cite{Arun:2006yw, Arun:2006hn, Yunes:2009ke, Mishra:2010tp}. In the TIGER framework~\cite{Agathos:2013upa, Li:2011cg, Meidam:2017dgf, Roy:2025gzv}, one or more post-Newtonian or phenomenological phase coefficients are allowed to differ from their GR predictions, providing a theory-agnostic test of the binary dynamics. GR corresponds to all deviation parameters being consistent with zero.
Schematically, a GR inspiral coefficient \(\varphi_k^{\mathrm{GR}}\) is modified as
\[
\varphi_k^{\mathrm{GR}}
\longrightarrow
\left(1+\delta\hat{\varphi}_k\right)
\varphi_k^{\mathrm{GR}},
\]where \(\delta\hat{\varphi}_k=0\) is the GR prediction. A statistically significant nonzero value could therefore indicate physics beyond GR. Such an interpretation, however, assumes that the GR waveform model accurately describes all relevant features of the source, and that the
detector noise is correctly described in the likelihood, typically as stationary and Gaussian with a known power spectral density (PSD). Non-Gaussian noise transients or PSD misestimation can themselves mimic a deviation~\cite{Kwok:2021zny, Gupta:2024gun}. In this work, we use zero-noise injections to isolate the effect of waveform inaccuracies. In this setting, any residual difference between the true signal and the recovery model can be absorbed by the additional degrees of freedom modeling GR deviations, producing an apparent violation of GR even when the true signal is fully consistent with Einstein's theory.

In particular, we focus on BBH signals. Current waveform approximants use different analytical descriptions of the inspiral, different treatments of spin precession, and different prescriptions for connecting the inspiral to the merger and ringdown. Comparisons with numerical relativity (NR) have shown that the accuracy of precessing waveform models can depend strongly on the mass ratio, spin configuration, inclination, and multipolar content~\cite{MacUilliam:2024oif}, and that waveform mismatches grow with signal length~\cite{Mitman:2025tmj}. Differences between waveform models can consequently lead to biased recovery of the component masses, spin magnitudes, effective-spin combinations, and spin orientations~\cite{Dhani:2024jja}.

More importantly, biases in the ordinary binary parameters may propagate into parameterized tests of GR. Previous studies have demonstrated that neglecting physical effects such as spin precession, higher-order modes, or orbital eccentricity can systematically shift non-GR parameters~\cite{Pang:2018hjb, Chandramouli:2024vhw, Roy:2025gzv, Narayan:2023vhm, Maggio:2022hre, Shaikh:2024wyn}. Astrophysical environmental effects~\cite{CanevaSantoro:2023aol, Roy:2024rhe} and gravitational lensing~\cite{Mishra:2023vzm, Wright:2024mco, Liu:2024xxn} can likewise produce apparent deviations; see Ref.~\cite{Gupta:2024gun} for a comprehensive overview. Even for quasi-circular binaries in vacuum GR, with spin precession and higher-order modes included in the waveform model, residual modelling errors can still mimic a violation: a recent injection--recovery study with 35 precessing NR simulations found that \phXPHM can produce apparent deviations from GR in the inspiral--merger--ringdown consistency test~\cite{Akcay:2025rve}. Mismodelling of the precessing ringdown can similarly bias this test~\cite{Foo:2024exr}. Additionally, waveform systematics have been shown to mimic lensing signatures during the recovery of simulated signals from very heavy BBH mergers like GW231123~\cite{Hegde:2026anm}.
 For sufficiently complex signals, waveform mismodelling can even in principle produce a statistical preference for a waveform containing an apparent non-GR correction.  These effects become increasingly important as the signal-to-noise ratio (SNR) increases because statistical uncertainties decrease while the underlying modelling error remains.

NR waveforms provide a natural benchmark for investigating these systematics because they are obtained by directly solving Einstein’s equations and do not rely on the approximate inspiral–merger–ringdown descriptions used in semi-analytical waveform models~\cite{Pretorius:2005gq, Campanelli:2005dd, Baker:2005vv}. NR simulations of highly spinning and precessing binaries are available from several groups; in particular, the moving-puncture approach has been used to evolve binaries with spins up to $\chi\simeq0.99$, with extensive public catalogs provided by the RIT group~\cite{Zlochower:2017bbg, Ficarra:2026cej}. For this study, we use the third SXS Collaboration catalog of BBH mergers~\cite{Scheel:2025jct}, which substantially expands the available coverage of precessing systems and includes simulations with dimensionless spin magnitudes up to approximately $0.8$, mass ratios up to $Q=8$ in its dense precessing coverage, and durations extending to 148 orbits. These long simulations provide a longer baseline for comparing the phase and precessional dynamics predicted by current waveform models against NR, and are particularly useful for parameterized inspiral tests, where longer NR waveforms reduce the need for hybridization with an analytical approximant~\cite{Hannam:2010ky, MacDonald:2011ne}.

In this work, we use long BBH simulations from the SXS catalog to investigate the impact of waveform systematics on parameterized inspiral tests of GR. We emphasize highly spinning systems to probe configurations with pronounced dynamical effects. The selected simulations span different spin magnitudes and orientations, including aligned, anti-aligned, and strongly precessing configurations. This allows us to examine whether the severity of waveform systematics is controlled primarily by the spin magnitude or by the detailed geometry and evolution of the binary.

We first perform NR injections and recover them under the assumption of GR using three state-of-the-art precessing waveform models: \phXPHMST~\cite{Colleoni:2024knd}, \phXPNR~\cite{Hamilton:2025xru}, and \SEOBvFive~\cite{Ramos-Buades:2023ehm}. These models provide complementary
descriptions of precessing-binary dynamics, as described in
Sec.~\ref{sec:waveform}.
The GR recovery study quantifies model-dependent biases in the masses, spin magnitudes, effective-spin parameters, and spin-orientation angles. Comparing several independent waveform families also helps distinguish effects that are specific to one approximant from those that are common across current precessing models. 

We then analyze the same NR injections using the \phXPHM -based TIGER framework, allowing individual inspiral phase coefficients to deviate from their GR values. This two-stage analysis enables us to determine how waveform inaccuracies first bias the inferred physical parameters and subsequently project onto the parameterized deformation directions. Although every injected waveform is a solution of GR, we find apparent nonzero deviations for a subset of the configurations considered. The effect is strongest for asymmetric binaries and grows with the SNR, so that a marginal shift at moderate SNR can become an apparently significant violation for loud signals.


The paper is organized as follows. In Sec.~\ref{sec:waveform}, we review the dynamics of spinning BBHs and the precessing waveform models used in this work, and describe the parameterized waveform within the TIGER framework and our Bayesian analysis setup. In Sec.~\ref{sec:RoleOfSpin}, we use controlled injections to study how the spin orientation and mass ratio affect the measurability of the PN deviation parameters, including the prior-dependent behaviour of nearly anti-aligned configurations. In Sec.~\ref{sec:NRInjections}, we analyze NR injections from the SXS catalog, first quantifying the biases in GR recovery with three waveform models, then studying the impact of these biases on the TIGER inspiral test, and finally investigating waveform systematics in the TIGER analysis of loud signals. We summarize our results and discuss their implications in Sec.~\ref{sec:conclusion}.

\section{Waveform model and parameterized test of general relativity}
\label{sec:waveform}

\subsection{Precessing gravitational-wave waveforms}

\begin{figure}[t]
  \centering%
\includegraphics[width=0.47\textwidth]{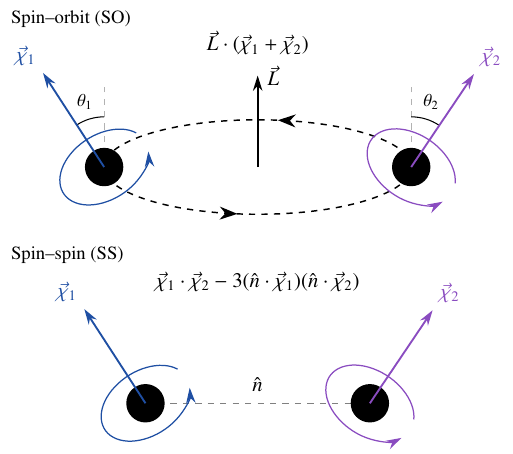}
  \caption{Schematic illustration of spin effects in BBH dynamics. The spin–orbit interaction couples the individual spins to the orbital angular momentum, while the spin–spin interaction depends on the relative orientation of the spins and the orbital separation.}
  \label{fig:schematicSpin}
\end{figure}

We describe a quasi-circular BBH by the component masses $m_1 \geq m_2$, with total mass $M = m_1 + m_2$, mass ratio $q = m_2/m_1 \leq 1$ (equivalently $Q = 1/q \geq 1$), and chirp mass
$\mathcal{M}_c = (m_1 m_2)^{3/5}/M^{1/5}$. The spins are described by the dimensionless spin vectors $\vec{\chi}_i = \vec{S}_i/m_i^2$, with magnitudes $\chi_i = |\vec{\chi}_i| \leq 1$ and tilt angles $\theta_i$ between $\vec{\chi}_i$ and the orbital angular momentum $\hat{L}$, defined at a reference frequency of $f_{\rm ref}$. The dominant aligned-spin effect on the inspiral is captured by the effective spin
\begin{equation}
\chi_{\rm eff} = \frac{m_1 \chi_1 \cos\theta_1 + m_2 \chi_2 \cos\theta_2}{M},
\end{equation}
while the in-plane spin components, $\chi_{i\perp} = \chi_i \sin\theta_i$, drive the precession of the orbital plane. These are commonly characterized by the effective precession parameter~\cite{Schmidt:2014iyl}
\begin{equation}
\chi_p = \max\left( \chi_{1\perp},\, \frac{4q+3}{4+3q}\, q\, \chi_{2\perp} \right),
\end{equation}
where $\chi_p = 0$ corresponds to a non-precessing binary.

The dynamics of spinning BBHs are governed by two main spin interactions, illustrated schematically in Fig.~\ref{fig:schematicSpin}. The spin--orbit interaction couples each BH spin to the orbital angular momentum and enters the GW phase at 1.5PN order, while the spin--spin interaction depends on the relative orientation of the two spins and the orbital separation and enters at 2PN order~\cite{Kidder:1992fr, Kidder:1995zr}. When the spins are aligned with $\hat{L}$, these interactions modify the rate of inspiral but leave the orbital plane fixed. For misaligned spins, the spin--orbit coupling causes $\vec{L}$ and the individual spins to precess around the total angular momentum $\vec{J}$~\cite{Apostolatos:1994mx}. Since precession occurs on a timescale longer than the orbital period but shorter than the inspiral time, the orbital plane changes orientation gradually throughout the inspiral.

Precession leaves characteristic imprints on the observed signal. As the orbital plane changes orientation relative to the line of sight, the GW amplitude and phase are modulated, and power is redistributed among the different $(\ell,m)$ multipoles, so that subdominant modes become more important, particularly for asymmetric and highly inclined systems. These effects carry additional information about the source, but they also make precessing signals more difficult to model accurately.

Current precessing waveform models typically describe the signal in a frame that co-precesses with the orbital plane, in which the waveform resembles that of an aligned-spin binary, and then rotate it to the inertial frame using time- or frequency-dependent Euler angles. This twisting-up approach~\cite{Schmidt:2010it, Boyle:2011gg, Schmidt:2012rh} is used, with different ingredients, by all three models employed in this work. \phXPHMST~\cite{Pratten:2020ceb, Colleoni:2024knd} combines the aligned-spin multipolar model \phXHM~\cite{Garcia-Quiros:2020qpx} with Euler angles obtained by numerically evolving the orbit-averaged PN spin-precession equations using \texttt{SpinTaylor} code~\cite{Sturani:2021SpinTaylor}. \phXPNR~\cite{Hamilton:2025xru} uses the same inspiral precession dynamics but calibrates the precession through merger and ringdown to NR simulations. \SEOBvFive~\cite{Ramos-Buades:2023ehm} instead models the full coalescence within the effective-one-body framework, evolving the spins with PN-expanded precession equations.

An important difference between the models is the treatment of asymmetries in the co-precessing-frame modes $h_{\ell m}$. In the \SEOBvFive analyses presented here, we enable the recently developed equatorial-asymmetric extension, which includes antisymmetric contributions to the $\ell\leq4$ modes~\cite{Estelles:2025zah}. \phXPNR also models the dominant co-precessing-frame multipole asymmetries, whereas \phXPHMST does not include these contributions.


\subsection{Parameterized waveforms in the TIGER approach}
Within the TIGER framework~\cite{Roy:2025gzv}, we construct the parameterized waveform model by incorporating a fractional deviation parameter ($\dpi_i$) into the phasing coefficients of phenomenological waveforms,
\begin{equation}
\varphi_i^{\mathrm{GR}} \rightarrow (1+\dpi_i)\: \varphi_i^{\mathrm{GR}},
\end{equation}
where $\varphi_i^{\mathrm{GR}}$ represents the \texttt{PhenomX} phasing coefficients~\cite{Pratten:2020fqn}, 
\begin{equation}
\label{eq:phiGR}
\varphi_i^{\mathrm{GR}} \equiv \big\{ \underbrace{ \varphi_0,...,\varphi_7,  \varphi_{5}^{(\ell)}, \varphi_{6}^{(\ell)}}_{\text{Inspiral}}, \  \underbrace{ b_1, b_2, b_3, b_4}_{\text{Intermediate}}, \ \underbrace{ c_1, c_2, c_4, c_L}_{\text{Merger-ringdown}}\big\}.
\end{equation}
For the spin-dependent inspiral coefficients $\varphi_3,\varphi_4,\varphi_5^{(\ell)},\varphi_6,$ and $\varphi_7$, the fractional deviation is applied only to the nonspinning contribution, while the spin-dependent contribution is unchanged. Many alternative theories of gravity include scalar fields alongside tensor fields, resulting in a leading dipolar contribution at -1PN and a tail-induced dipole contribution at 0.5PN~\cite{Will:1994fb, Yunes:2011aa, Arun:2012hf, Barausse:2016eii, Yagi:2015oca}. In contrast, these terms are absent in GR. To search for the non-GR dipolar radiation, we introduce the absolute deviation parameters at -1PN and 0.5PN. 


The higher-order modes of GW signals from coalescing compact binaries play a vital role in the accurate reconstruction of source properties. Including higher-order modes improves the measurement of the mass parameters, and since the deviation parameters are strongly correlated with $\mathcal{M}_c$, this also tightens the constraints on them. In contrast, the absence of higher-order modes in the parameterized framework can lead to false deviations of GR if their contribution is large in the signal~\cite{Pang:2018hjb}. The TIGER framework used here mitigates this issue by propagating the inspiral phase deviations to the higher-order modes~\cite{Roy:2025gzv}, as described below.

For non-precessing binaries, the inspiral phase of an arbitrary $(\ell, m)$ mode can be expressed in terms of the phase of the $(2,2)$ mode alone: ${ \Phi_{\ell m}(t) \simeq   (m/2) \; \Phi_{22}(t) }$. In the Fourier domain, this becomes, $\Phi^{\mathrm{Ins}}_{\ell m}(f) \approx (m/2) \, \Phi^{\mathrm{Ins}}_{22}\left( 2f/m \right)$. At a given PN order, if a phase correction is required for the $(2, 2)$ mode to accurately represent the signal, the same correction must be included in the $(\ell, m)$ mode phase, scaled by a factor of $m/2$. Using the scaling relation, the non-GR correction to the inspiral phase for $(\ell, m)$ mode is expressed as,
\begin{align}
\label{eq:dphilm}
\Delta\Phi_{\ell m} &= \frac{m}2{} \frac{3}{128\eta} \left(\frac{2\pi M f}{m}\right)^{(k-5)/3} \times \nonumber \\
&\begin{cases}
 \delta\hat{\varphi}_{k}^{(\ell)} \: \varphi_{k}^{(\ell)}(\vec{\lambda}) \: \log{\left(\frac{2\pi M f}{m}\right)}, & \text{for log terms}, \\
 \delta\hat{\varphi}_k , & \text{for $k=-2, 1$}, \\
\delta\hat{\varphi}_k \: \varphi_k(\vec{\lambda}), & \text{otherwise}
\end{cases}
\end{align}
where the superscript $(\ell)$ denotes the coefficients of the log terms in the PN expansion, not related to the $(\ell, m)$ mode.

The scaling relation between the modes is only valid during the inspiral phase and does not extend to the intermediate and merger-ringdown stages. In the \phXHM model~\cite{Garcia-Quiros:2020qpx}, the phenomenological coefficients for the post-inspiral part are modeled separately for individual modes. As a result, deviations in the post-inspiral phase do not affect the higher-order modes. No post-inspiral deviation parameters are introduced for the higher-order modes in the TIGER framework.

For precessing binaries, the parametrized phase deviations are introduced in the co-precessing frame modes and subsequently rotated to the inertial frame using the same twisting-up prescription as the GR waveform~\cite{Roy:2025gzv}. The precession dynamics themselves are not modified by the deviation parameters, so the Euler angles are determined from the GR binary evolution. Throughout this work, we use \phXPHMST~\cite{Colleoni:2024knd} as the baseline model for all TIGER analyses. In this model, the twisting-up angles are obtained using the numerical \texttt{SpinTaylor} precession prescription~\cite{Sturani:2021SpinTaylor}.

\subsection{Bayesian analysis framework}
The parameterized test with the \phX family described above introduces a set of non-GR parameters,
\begin{align}
\dpi_i \equiv & \Big\{\overbrace{ \delta\hat{\varphi}_{-2}, \delta\hat{\varphi}_0,..., \delta\hat{\varphi}_7,    \delta\hat{\varphi}_{5}^{(\ell)}, \delta\hat{\varphi}_{6}^{(\ell)}}^{\text{\normalsize Inspiral}},    \quad \overbrace{ \delta\hat{b}_1, \delta\hat{b}_2, \delta\hat{b}_3, \delta\hat{b}_4}^{\text{\normalsize Intermediate}}, \nonumber \\
& \quad \quad \quad \quad \quad \quad  \underbrace{\delta\hat{c}_1, \delta\hat{c}_2, \delta\hat{c}_4, \delta\hat{c}_L}_{\text{\normalsize Merger-ringdown}}\Big\},
\end{align}
which are applied to the \texttt{PhenomX} phasing coefficients as defined in Eq.~\eqref{eq:phiGR} and 2 additional parameters for deviation at -1PN and 0.5PN as explained above. Our parameterized test involves a nested hypothesis where GR is a special case. The beyond-GR model is characterized by all the standard GR signal parameters ($\vec{\theta}_{\mathrm{GR}}$) plus one additional non-GR parameter \mbox{$\dpi_i$, such that the signal parameter space is $\vec{\theta} \equiv \{\vec{\theta}_{\mathrm{GR}}, \dpi_i \}$}. For a given data $d$ and hypothesis $\mathcal{H}$, we obtain the posterior probability density \mbox{$p\big(\vec{\theta} \mid d,\mathcal{H}\big)$} using Bayes' theorem,
\begin{equation}
\label{eq:posterior}
p\big(\vec{\theta} \mid d, \mathcal{H} \big) \propto p\big(d \mid \vec{\theta}, \mathcal{H} \big) \; p\big(\vec{\theta} \mid \mathcal{H} \big),
\end{equation}
where \mbox{$p\big(\vec{\theta} \mid \mathcal{H} \big)$} is the prior probability distribution, and \mbox{$p\big(d \mid \vec{\theta}, \mathcal{H} \big)$} is the likelihood. We use the standard Whittle likelihood function assuming additive noise that is stationary and Gaussian~\cite{Veitch:2009hd},
\begin{equation}
p\big(d \mid \vec{\theta}, \mathcal{H} \big) \propto \exp\left[ -\frac{1}{2} \left< d - h(\vec{\theta}) \mid d - h(\vec{\theta}) \right>  \right],
\end{equation}
where the angular brackets, $\inp{\cdot}{\cdot}$, represent the noise-weighted inner product, defined as
\begin{equation}
 \inp{a}{b} \equiv 4 \Re \int_{\flow}^{\fhigh}  df \, \frac{ \tilde{a}^{\ast}(f) \, \tilde{b}(f) }{S_n (f)} \,, 
\end{equation}
and $S_n(f)$ is the one-sided PSD of the detector noise, and $ \tilde{a}(f) $ and $\tilde{b}(f)$ denote the Fourier transform of two time-series signals $a(t)$ and $b(t)$, respectively. Throughout this work, we use the advanced-design PSD \texttt{aLIGOZeroDetHighPower}~\cite{aLIGO_ZDHP} and the single-detector configuration with LIGO-Hanford (H1). We adopt this setup for simplicity, since the analyses are performed in zero noise at fixed optimal SNR, and the main impact of using a detector network would be on the extrinsic parameters, which are not the focus of this study. The integration limits \flow and \fhigh represent the lowest and highest cutoff frequencies of the sensitivity bandwidth. We perform all the analyses with a lower cutoff frequency of 20\Hz. With this setup, we fix the injection optimal SNR at 30 for all the analyses, except for the loud injection case in Sec.~\ref{subsec:loud}.

For the prior probability distribution \mbox{$p\big(\vec{\theta} \mid \mathcal{H} \big)$}, we adopt the setup used in the tests of GR published by the LIGO-Virgo-KAGRA collaboration~\cite{LIGOScientific:2026fcf}. For the GR parameters, we assume a uniform prior on component masses, a uniform prior on individual spin magnitudes with isotropic orientations, and a luminosity distance $d_L$ prior that is uniform in comoving volume, given by $p(d_L) \propto d_L^2$. For the non-GR parameter, we assume a uniform prior symmetric around zero.

We estimate the beyond-GR effects by computing the marginalized posterior probability distribution for $\dpi_i$, 
\begin{equation}
\label{eq:dpi_posterior}
    p\left(\dpi_i \,|\, d \right) = \int \left( \prod_{\vec{\theta} \,\setminus \{\dpi_i\}} d\theta_i \right) \: p( \vec{\theta} \,|\, d,  \mathcal{H} ).
\end{equation}
To quantify the consistency of the recovered deviation parameter with GR, we compute the GR quantile, $Q_{\rm GR}$, defined as the fraction of the posterior enclosed by the smallest highest-posterior-density interval containing the GR value. Smaller values of $Q_{\rm GR}$ indicate greater consistency with GR, while values approaching $100\%$ correspond to increasing tension with the GR prediction.

Throughout this work, we perform Bayesian inference using the \bilbytgr package~\cite{bilby-tgr} based on \bilby~\citep{Ashton:2018jfp}, with the \dynesty nested sampling sampler~\cite{Speagle:2019ivv}. We use 1000 live points, the \texttt{acceptance-walk} sampling method, and set the acceptance parameter to \texttt{naccept}=60\footnote{Further details on the \dynesty sampler settings are available at \url{https://bilby-dev.github.io/bilby/dynesty-guide.html}.}.

\section{Role of spin in parameterized tests of general relativity}
\label{sec:RoleOfSpin}
\begin{figure*}[htbp]
\centering
  \includegraphics[width=0.99\textwidth]{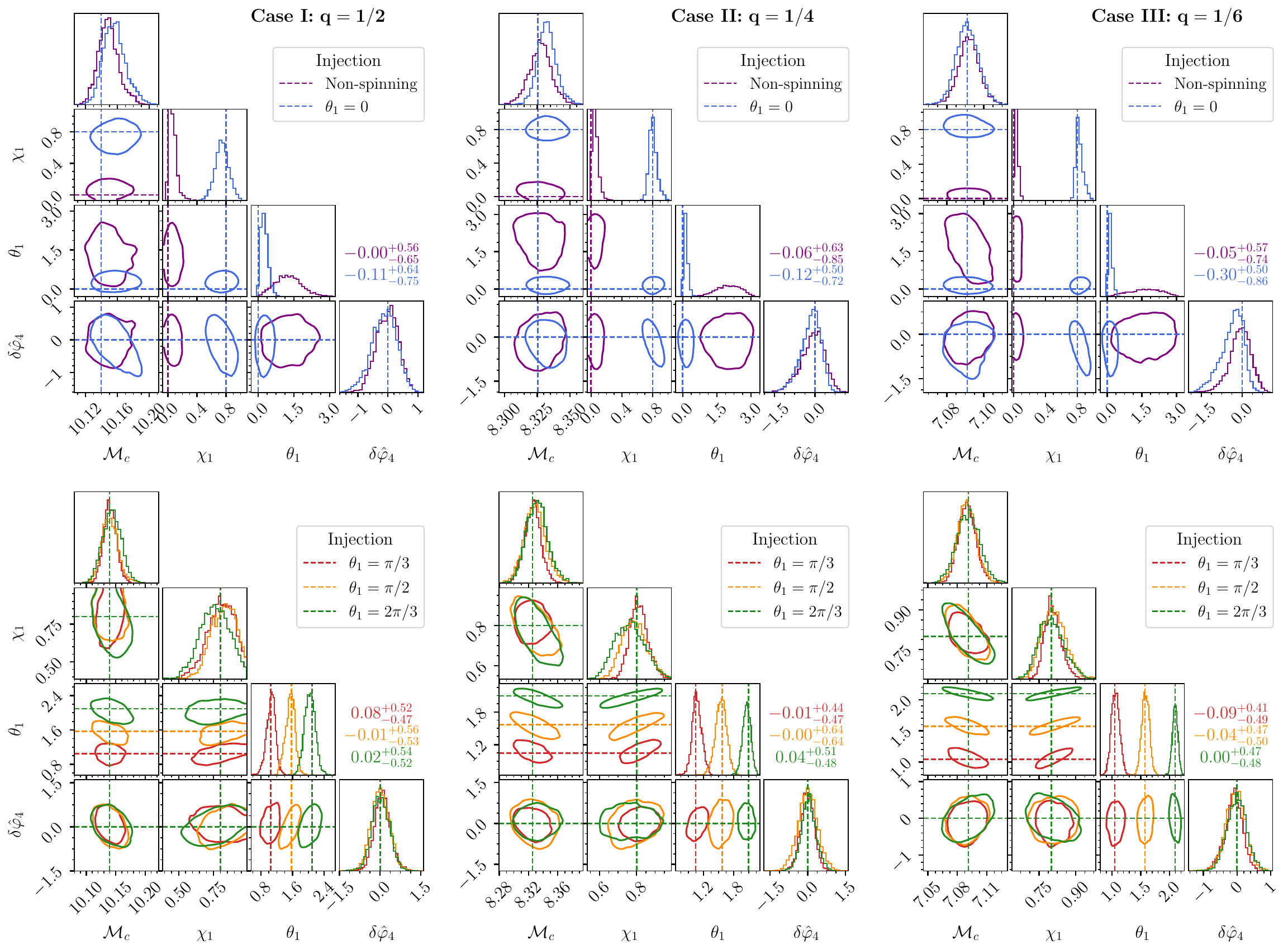} 
  \caption{Illustrating the measurability of parameterized deviations from GR as a function of spin orientation and binary mass asymmetry. Shown are the joint and marginalized posterior distributions of the primary-spin magnitude $\chi_1$, tilt angle $\theta_1$, chirp mass $\mathcal{M}_c$, and the 2PN deviation parameter $\delta\hat{\varphi}_4$ obtained from TIGER analyses of GR injections with mass ratios $q=1/2$, $1/4$, and $1/6$ (columns). The top row shows the nonspinning and aligned-spin injections, while the bottom row shows precessing injections with primary-spin tilt angles $\theta_1=\pi/3$, $\pi/2$, and $2\pi/3$. The secondary BH is nonspinning in all cases. The dash-dotted lines indicate the injected values, and the values above the $\delta\hat{\varphi}_4$ posteriors show the corresponding 90\% credible intervals.}
  \label{fig:SpinInjsTIGER}
\end{figure*}

Higher-PN deviations remain comparatively poorly constrained by current GW observations~\cite{LIGOScientific:2026fcf, LIGOScientific:2026oim}. In spinning binaries, higher-PN terms contain increasingly important spin-dependent contributions and become more prominent toward the late inspiral as the orbital velocity increases. We therefore use the 2PN coefficient $\delta\hat{\varphi}_4$ as our primary case study, where spin-spin effects first enter the GR phasing.

To understand the role of spin orientation in parameterized tests of GR, we perform a series of controlled injections in which the spin magnitude of the primary BH is fixed to \mbox{$\chi_1 = 0.8$} while its tilt angle $\theta_1$ is varied. The secondary BH is assumed to be nonspinning. We consider a binary total mass of $M=25\:\msun$ and three mass ratios, $q=1/2$, $1/4$, and $1/6$, and compare each spinning configuration with the corresponding nonspinning binary. For the spinning systems, we consider tilt angles $\theta_1 = 0$, $\pi/3$,
$\pi/2$, $2\pi/3$, and $\pi$, corresponding to aligned, precessing, and
anti-aligned primary spins. All injections are generated in zero noise using \phXPHMST, with an inclination of $\iota=\pi/3$, a reference frequency of $\fref=20\,\Hz$, and an optimal SNR of 30. The same waveform model is used for the recovery, so waveform-systematic effects are absent by construction, and any shifts in the recovered posteriors arise from parameter correlations or prior effects.

The correlations discussed below are therefore specific to this coefficient. However, the prior-dependent effect for the anti-aligned configurations, discussed in Sec.~\ref{subsec:antialigned}, arises from the recovery of the GR parameters and and is therefore distinct from the PN-order dependent effect of spin orientation on the measurability of the deviation parameters. In Sec.~\ref{subsec:otherPN}, we further investigate the $-1$PN and 3PN coefficients and find that the impact of spin orientation depends on the PN order.

\subsection{Non-spinning and aligned-spin binaries}
We first discuss the non-spinning and aligned-spin configurations to understand correlations with the 2PN deviation parameter $\delta\hat{\varphi}_4$. Results are shown in the top row of Fig.~\ref{fig:SpinInjsTIGER}. For the nonspinning binaries, $\delta\hat{\varphi}_4$ shows a strong correlation with the chirp mass $\mathcal{M}_c$, particularly for the less asymmetric system with $q=1/2$. In this case, a change in the 2PN phase coefficient can be partially compensated by a shift in $\mathcal{M}_c$, producing an elongated joint posterior in the $\mathcal{M}_c$--$\delta\hat{\varphi}_4$ plane. As the binary becomes more asymmetric, from $q=1/2$ to $q=1/6$, this correlation becomes progressively weaker, and the two parameters are more clearly separated. This indicates that increasing mass asymmetry helps to break the degeneracy between the chirp mass and the parameterized 2PN correction.

The situation is different for the aligned-spin configurations with $\theta_1=0$. Once the primary BH carries a large aligned spin, the dominant correlation of $\delta\hat{\varphi}_4$ shifts from $\mathcal{M}_c$ to the spin magnitude $\chi_1$. The $\chi_1$--$\delta\hat{\varphi}_4$ contours show a clear degeneracy in which larger values of $\chi_1$ are associated with more negative values of $\delta\hat{\varphi}_4$. Unlike the $\mathcal{M}_c$--$\delta\hat{\varphi}_4$ degeneracy in the non-spinning case, this correlation remains prominent as the mass ratio decreases from $q=1/2$ to $q=1/6$. Therefore, increasing the mass asymmetry alone does not remove the degeneracy between the aligned spin and the higher-PN deviation parameter.

The marginalized posteriors of $\delta\hat{\varphi}_4$ are consistent with the injected GR value for all three non-spinning systems. We obtain
$\delta\hat{\varphi}_4=-0.00^{+0.56}_{-0.65}$,
$-0.06^{+0.63}_{-0.85}$, and
$-0.05^{+0.57}_{-0.74}$
for $q=1/2$, $1/4$, and $1/6$, respectively.
Although the correlation with $\mathcal{M}_c$ becomes weaker for more asymmetric binaries, the marginalized uncertainty on $\delta\hat{\varphi}_4$ does not improve substantially. This shows that removing a particular parameter degeneracy does not necessarily translate directly into a much tighter one-dimensional constraint on the deviation parameter.

For the aligned-spin injections, the recovered values are
$\delta\hat{\varphi}_4=-0.11^{+0.64}_{-0.75}$,
$-0.12^{+0.50}_{-0.72}$, and
$-0.30^{+0.50}_{-0.86}$
for $q=1/2$, $1/4$, and $1/6$, respectively.
The posteriors remain consistent with GR, although their medians progressively shift towards negative values as the binary becomes more asymmetric. This shift follows the $\chi_1$--$\delta\hat{\varphi}_4$ degeneracy seen in the joint posteriors: the recovery favours slightly larger values of the aligned spin together with a negative 2PN deviation. The persistence of this correlation across the three mass ratios suggests that, for highly spinning aligned binaries, part of the higher-PN phasing information can remain degenerate with the spin contribution even in strongly asymmetric systems.

\subsection{Precessing binaries}
We next discuss the precessing configurations with $\theta_1=\pi/3$, $\pi/2$, and $2\pi/3$, as shown in the bottom row of Fig.~\ref{fig:SpinInjsTIGER}. In contrast to the aligned-spin case, the 2PN deviation parameter $\delta\hat{\varphi}_4$ shows very little correlation with the GR parameters once the primary spin is tilted away from the orbital angular momentum. For $q=1/2$, a mild correlation between $\delta\hat{\varphi}_4$ and the tilt angle $\theta_1$ is still visible for the $\theta_1=\pi/2$ and $2\pi/3$ injections. However, this correlation becomes much weaker for $q=1/4$ and is almost absent for $q=1/6$. This behaviour indicates that the additional information introduced by spin precession helps to separate the parameterized phase correction from the intrinsic spin degrees of freedom, with the separation becoming clearer for more asymmetric binaries.

The marginalized $90\%$ credible intervals show a similar trend. For $q=1/2$, we obtain
$\delta\hat{\varphi}_4=0.08^{+0.52}_{-0.47}$,
$-0.01^{+0.56}_{-0.53}$, and
$0.02^{+0.54}_{-0.52}$
for $\theta_1=\pi/3$, $\pi/2$, and $2\pi/3$, respectively.
For $q=1/4$, the corresponding constraints are
$-0.01^{+0.44}_{-0.47}$,
$-0.00^{+0.64}_{-0.64}$, and
$0.04^{+0.51}_{-0.48}$,
while for $q=1/6$ they become
$-0.09^{+0.41}_{-0.49}$,
$-0.04^{+0.47}_{-0.50}$, and
$0.00^{+0.47}_{-0.48}$.
All of these measurements remain consistent with the injected GR value. 
Among the three precessing configurations, the $\theta_1=\pi/3$ case gives the tightest constraint on $\delta\hat{\varphi}_4$ for each of the three mass ratios. The improvement is particularly clear for the more asymmetric systems, where the $\delta\hat{\varphi}_4$ posterior is both well centered around the GR value and only weakly correlated with the GR parameters.

From the waveform perspective, $\delta\hat{\varphi}_4$ and an aligned spin both enter primarily through the phase and can therefore partially compensate each other, whereas an in-plane spin additionally induces precession-driven amplitude and phase modulations that a phase deformation cannot reproduce. These modulations constrain the spin independently of $\delta\hat{\varphi}_4$, and become stronger for asymmetric binaries, where the orbital angular momentum is smaller relative to the primary spin. The $\theta_1=\pi/3$ configuration additionally retains a positive spin projection onto $\hat{L}$, which increases the number of inspiral cycles in band.

\begin{figure*}[t!]
\centering
  \includegraphics[width=0.97\textwidth]{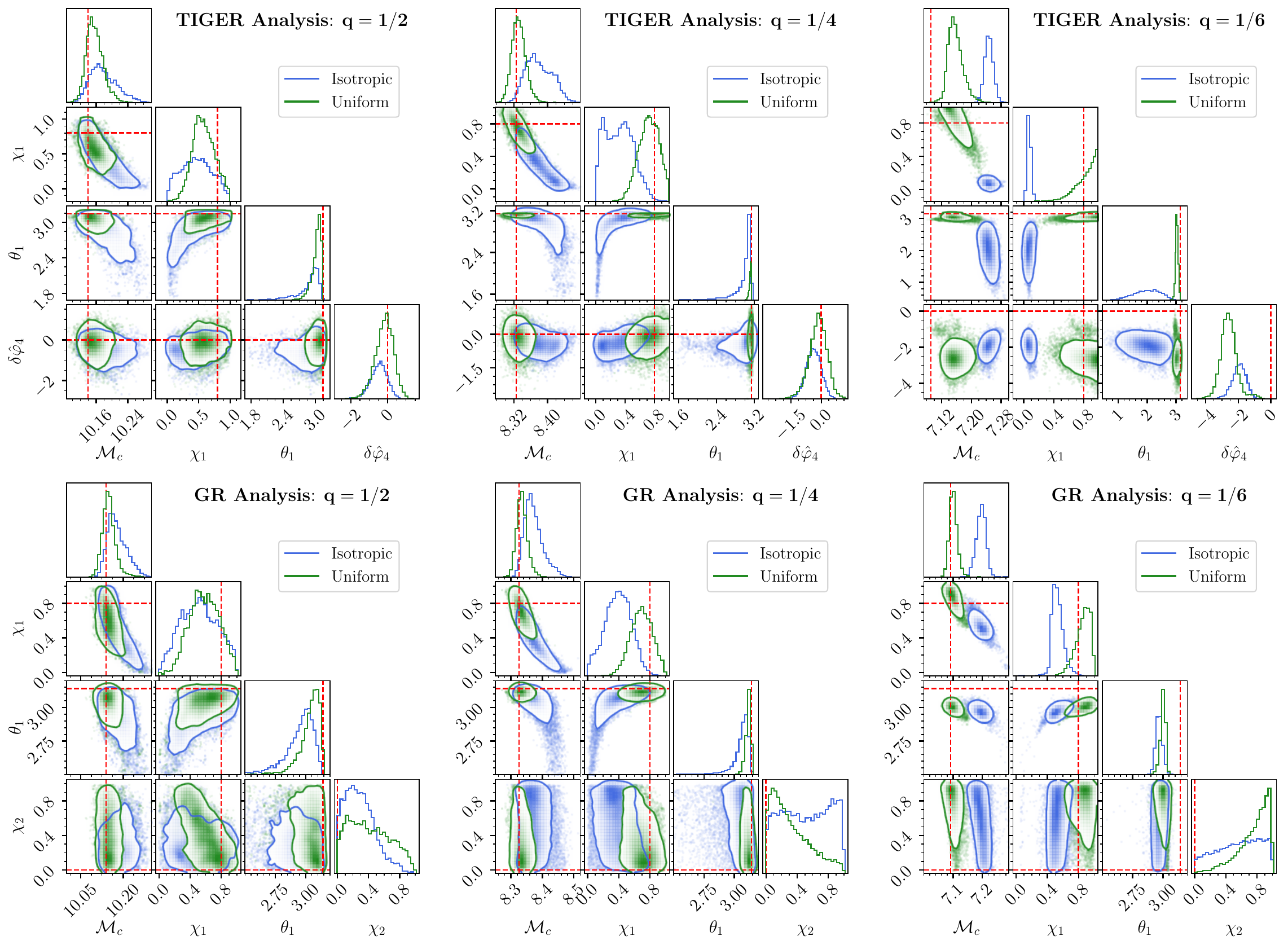}
  \caption{Diagnosing the origin of the biases observed for anti-aligned-spin injections. Shown are the joint and marginalized posterior distributions obtained using the standard isotropic spin-orientation prior, $p(\theta_1)\propto\sin\theta_1$ (blue), and a prior uniform in the primary-spin tilt angle $\theta_1$ (green). The top row shows the TIGER analyses for mass ratios $q=1/2$, $1/4$, and $1/6$, including the chirp mass $\mathcal{M}_c$, primary-spin magnitude $\chi_1$, tilt angle $\theta_1$, and the 2PN deviation parameter $\delta\hat{\varphi}_4$. The bottom row shows the corresponding GR-only recoveries, used to assess whether the shifts in the intrinsic parameters persist in the absence of a parameterized deviation. The red dashed lines indicate the injected values. For $q=1/2$ and $q=1/4$, adopting a prior uniform in $\theta_1$ largely restores the injected parameters and brings $\dphi_4$ closer to zero, whereas for $q=1/6$ the TIGER analysis still excludes the GR value.}
  \label{fig:antialigned_prior}
\end{figure*}

Although the degeneracy between the TIGER parameter and the GR parameters is substantially reduced, correlations among the GR parameters themselves remain present. For $q=1/4$ and $q=1/6$, the joint posteriors show clear correlations between $\mathcal{M}_c$ and $\theta_1$, as well as between $\chi_1$ and $\theta_1$. These correlations reflect that different combinations of the spin magnitude and orientation can produce similar precessional dynamics while also correlating with the mass parameters. The same structure is less clearly visible for $q=1/2$. In this case, however, the primary spin magnitude is comparatively poorly constrained, with the posterior extending towards the physical upper boundary $\chi_1=1$, which makes the underlying $\chi_1$--$\theta_1$ correlation less distinct.

\begin{figure*}[htbp]
\centering  \includegraphics[width=0.98\textwidth]{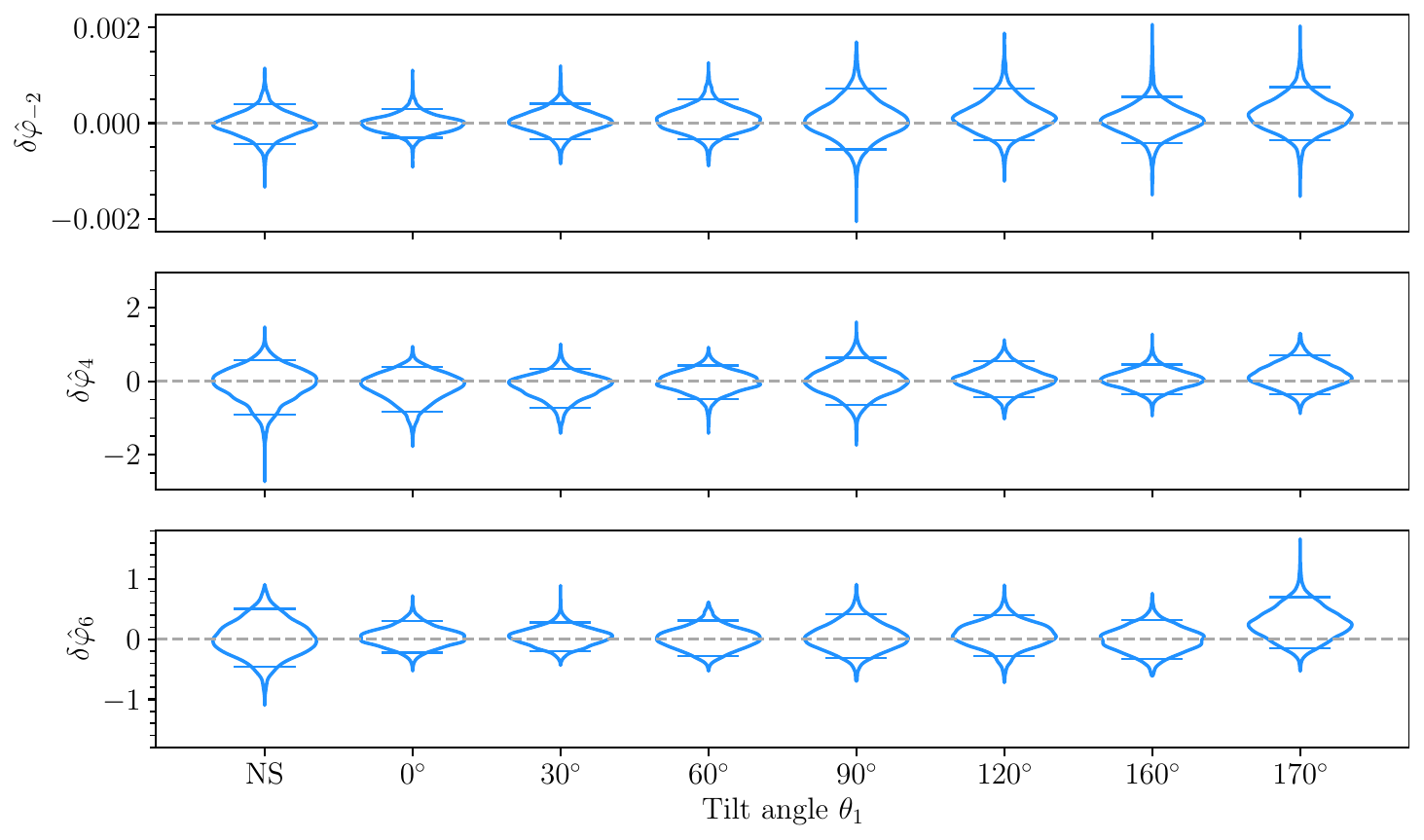}
  \caption{Illustrating the measurability of parameterized deviations from GR as a function of the primary-spin orientation for a binary with mass ratio $q=1/4$. Shown are the marginalized posterior distributions of the $-1$PN deviation parameter $\delta\hat{\varphi}_{-2}$ (top), the 2PN deviation parameter $\delta\hat{\varphi}_{4}$ (middle), and the 3PN deviation parameter $\delta\hat{\varphi}_{6}$ (bottom) for a nonspinning (NS) injection and spinning injections with different primary-spin orientations, while the secondary BH is nonspinning. The horizontal dashed lines denote the injected GR value, $\delta\hat{\varphi}_i=0$, and the horizontal bars indicate the corresponding $90\%$ credible intervals.}
  \label{fig:SpinTiltViolinTIGER}
\end{figure*}
Overall, the precessing injections show that spin misalignment can substantially reduce the degeneracy between the higher-PN deviation parameter and the intrinsic GR parameters. This is qualitatively different from the aligned-spin configurations, where $\delta\hat{\varphi}_4$ remains strongly correlated with $\chi_1$. The effect becomes more apparent as the binary becomes more asymmetric, suggesting that precessing, asymmetric BBHs provide a particularly favorable configuration for constraining deviations entering at higher PN orders.

\subsection{Anti-aligned spin binaries}
\label{subsec:antialigned}
We finally consider anti-aligned configurations. In the initial TIGER analyses, these systems exhibit substantial shifts in both the intrinsic source parameters and the 2PN deviation parameter $\delta\hat{\varphi}_4$, with the effect becoming particularly pronounced for the most asymmetric binaries, as shown in Fig.~\ref{fig:antialigned_prior}. At first sight, this behaviour may suggest a strong degeneracy between the parameterized deviation and the intrinsic GR parameters. However, additional tests show that this interpretation is incomplete.

To identify the origin of the bias, we perform corresponding GR-only recoveries, without allowing for any parameterized deviation. We find similar shifts in the recovered masses and spins, including a substantial underestimate of the primary-spin magnitude for the anti-aligned injections. Moreover, this behaviour is reproduced across several waveform models, including \SEOBvFive, \phXPHMST, and alternative \phXPHM prescriptions. The persistence of the bias across both TIGER and GR-only analyses, and across different waveform families, indicates that it is not primarily caused by the introduction of the parameterized deviation or by waveform-model systematics.

We also repeat the analysis with a more stringent sampler configuration, using $N_{\rm live}=2000$, $\texttt{maxmcmc}=15000$, and $\texttt{naccept}=100$, and find no noticeable improvement in the recovered posteriors. The recovered posteriors are stable under these changes to the sampler settings, providing no evidence that the shifts are primarily caused by insufficient sampling.

We further verify that the waveform overlap varies smoothly as the primary-spin tilt is moved away from the anti-aligned configuration, with the match decreasing gradually from $\theta_1=\pi$, indicating no apparent discontinuity or pathological behaviour of the waveform model in this region.

We instead find that the behaviour is strongly sensitive to the prior on the spin orientation. Under the commonly adopted isotropic spin-orientation prior, the tilt distribution satisfies $p(\theta_1)\propto\sin\theta_1$, and therefore the prior density vanishes at the exactly aligned and anti-aligned configurations, $\theta_1=0$ and $\pi$. Consequently, an exactly anti-aligned injection lies at a region of very small prior volume. Near $\theta_1=\pi$, the available prior volume decreases rapidly, allowing posterior-volume effects to favour configurations displaced from the injected anti-aligned solution even when the likelihood remains compatible with it.

To test this interpretation, we repeat the analysis using a prior uniform in $\theta_1$. Fig.~\ref{fig:antialigned_prior} compares the resulting posteriors with those obtained using the isotropic prior, for the TIGER (top row) and GR-only (bottom row) analyses. For $q=1/2$, the prior has only a small effect on the GR parameters, but the uniform-tilt prior brings the $\dphi_4$ posterior back to zero. For $q=1/4$, the uniform-tilt prior brings the intrinsic parameters closer to their injected values in both analyses and reduces the shift in $\delta\hat\varphi_4$. This indicates that the apparent deviation is strongly sensitive to the spin-orientation prior.

The most asymmetric system, $q=1/6$, behaves differently. In the GR-only analysis, the uniform-tilt prior restores $\mathcal{M}_c$ and $\chi_1$ close to their injected values, with small residual offsets in $\theta_1$ and a spurious secondary spin. In the TIGER analysis, however, the uniform-tilt prior recovers the injected tilt angle, but $\dphi_4$ moves further away from zero, together with an overestimated $\mathcal{M}_c$ and $\chi_1$ pushed towards the upper prior boundary. The displacement in $\mathcal{M}_c$ and $\chi_1$ is substantially reduced in the GR-only recovery under the same prior but remains in the TIGER analysis. Together with the joint posteriors, this behaviour is consistent with a degeneracy between $\delta\hat\varphi_4$, $\mathcal{M}_c$, and $\chi_1$.

These results show that the apparent deviations in the tested anti-aligned configurations reflect an interplay between prior effects and parameter degeneracies. For $q=1/2$ and $q=1/4$, adopting a uniform-tilt prior reduces the displacement of $\delta\hat\varphi_4$, indicating a substantial prior-dependent contribution. For $q=1/6$, however, the GR-only recovery improves while the TIGER posterior remains displaced from zero. This behaviour is consistent with an additional degeneracy between the deformation parameter and the intrinsic parameters. Apparent deviations can therefore arise in these configurations even when the injection and recovery waveform models are identical.

\subsection{Role of spin at -1PN and 3PN}
\label{subsec:otherPN}
We next investigate whether the spin-orientation dependence found for the 2PN deviation persists at other PN orders. Fig.~\ref{fig:SpinTiltViolinTIGER} shows the marginalized posteriors of the $-1$PN and 3PN deviation parameters, $\delta\hat{\varphi}_{-2}$ and $\delta\hat{\varphi}_{6}$, for the same $q=1/4$ injections with different primary-spin orientations. For both parameters, the constraints depend noticeably on the spin orientation. The positively aligned configurations, with $\theta_1<90^\circ$, generally provide tighter constraints than configurations with a negative projected spin, $\theta_1>90^\circ$. This behaviour is particularly clear for the $-1$PN deviation, where the $90\%$ credible interval is narrowest for the aligned and mildly tilted configurations and broadens as the spin is tilted towards the anti-aligned direction.

A similar trend is observed for the 3PN coefficient. The $90\%$ credible intervals are comparatively narrow for $\theta_1=0^\circ$, $30^\circ$, and $60^\circ$, while broader posteriors are obtained for most configurations with $\theta_1>90^\circ$. Quantitatively, the characteristic width of the $90\%$ interval increases by roughly $40\%$ for $\delta\hat{\varphi}_{-2}$ and by about $35$--$40\%$ for $\delta\hat{\varphi}_{6}$ when comparing the positive- and negative-projected-spin configurations. In contrast, the 2PN deviation does not show such a clear monotonic asymmetry, indicating that the impact of spin orientation on the measurability of parameterized deviations depends on the PN order. Nevertheless, the spinning configurations generally provide tighter constraints than the nonspinning case, indicating that spin can still improve the measurability of the 2PN deviation even though the dependence on its orientation is more complex.

These results suggest that the spin contribution can either help or hinder the measurement of non-GR phase corrections depending on both its orientation and the PN order at which the deviation enters. In particular, positive projected spins appear to improve the measurability of the $-1$PN and 3PN coefficients, whereas the behaviour at 2PN is more strongly affected by parameter correlations. We do not use the nearly anti-aligned configurations to infer this trend, since their posteriors are additionally affected by the prior-volume effects discussed in the previous subsection.

\section{Impact of waveform systematics on parameterized tests of general relativity}
\label{sec:NRInjections}
\begin{table}[t!]
    \centering
    \renewcommand{\arraystretch}{1.5}
    \setlength{\tabcolsep}{6pt}
    \begin{tabular}{lcccccc}
        \toprule[1pt]
        \toprule[1pt]
        \textsc{SXS ID}
        & $Q$
        & $\abs{\vec{\chi}_1}$
        & $\abs{\vec{\chi}_2}$
        & $\chi_{\rm eff}$
        & $\chi_{\rm p}$
        & $N_{\rm orb}$ \\
        \midrule[0.5pt]

        \rowcolor{rowshade}[\tabcolsep]
        \sxsid{1400} & 1.88 & 0.50 & 0.50 & $-0.12$ & 0.38 & 63.5 \\
        \sxsid{1408} & 1.67 & 0.46 & 0.38 & $-0.17$ & 0.28 & 61.2 \\
        
        \rowcolor{rowshade}[\tabcolsep]
        \sxsid{2322} & 1.11 & 0.80 & 0.80 & $-0.79$ & 0.14 & 94.6 \\
        \sxsid{2619} & 1.83 & 0.47 & 0.44 & $-0.23$ & 0.33 & 143.1 \\
        
        \rowcolor{rowshade}[\tabcolsep]
        \sxsid{2622} & 1.46 & 0.41 & 0.50 & $-0.23$ & 0.32 & 59.1 \\
        \sxsid{2623} & 1.91 & 0.47 & 0.47 & 0.26 & 0.40 & 68.6 \\
        
        \rowcolor{rowshade}[\tabcolsep]
        \sxsid{2625} & 4.00 & 0.50 & 0.47 & $-0.16$ & 0.40 & 69.9 \\
        \sxsid{2626} & 4.00 & 0.50 & 0.49 & 0.17 & 0.45 & 76.1 \\
        
        \rowcolor{rowshade}[\tabcolsep]
        \sxsid{2632} & 4.00 & 0.80 & 0.00 & 0.46 & 0.56 & 73.3 \\
        \sxsid{2634} & 4.00 & 0.80 & 0.00 & $-0.45$ & 0.57 & 64.4 \\
        
        \rowcolor{rowshade}[\tabcolsep]
        \sxsid{2633} & 8.00 & 0.80 & 0.00 & 0.51 & 0.56 & 86.7 \\
        \sxsid{2635} & 8.00 & 0.80 & 0.00 & $-0.50$ & 0.57 & 72.2 \\

        \bottomrule[1pt]
        \bottomrule[1pt]
    \end{tabular}
    \caption{
    Properties of the selected SXS numerical-relativity simulations.
    Here $Q=m_1/m_2\geq1$ is the mass ratio,
    $\abs{\vec{\chi}_1}$ and $\abs{\vec{\chi}_2}$ are the dimensionless spin magnitudes,
    $\chi_{\rm eff}$ is the effective aligned spin,
    $\chi_{\rm p}$ is the effective precession spin,
    and $N_{\rm orb}$ denotes the number of orbits in the simulation.
    }
    \label{tab:sxs_simulations}
\end{table}

To assess the impact of waveform systematics on parameterized tests of GR, we consider a set of long-duration SXS simulations of BBH systems, listed in
Table~\ref{tab:sxs_simulations}. These simulations were chosen to cover
mass ratios from nearly equal mass to moderately high asymmetry
($Q = 1/q \simeq 1$--$8$) and moderated to high spin magnitudes ($|\vec{\chi}_i| \simeq 0.4$--$0.8$). Using these NR waveforms, we construct injection signals in a zero-noise realization to isolate waveform-systematic effects from noise artefacts. We retain NR modes with angular multipole index $\ell\leq4$ when constructing the injections. The NR modes contain the same number of cycles, implying different starting frequencies for different $m$-modes. We therefore scale each simulation to a total mass such that the $(4,4)$ mode starts at $20\Hz$, corresponding to a starting frequency of $10\Hz$ for the dominant $(2,2)$ mode\footnote{Analysis recovery is still done starting at 20Hz.}. We set the inclination angle to $\theta_{\rm JN}=60^\circ$ and the optimal injection SNR to 30. We discuss the injection and recovery of six NR simulations here and show results for the remaining ones in Appendix~\ref{app:GR}.

\begin{figure*}[htbp]
\centering
  \includegraphics[width=0.97\textwidth]{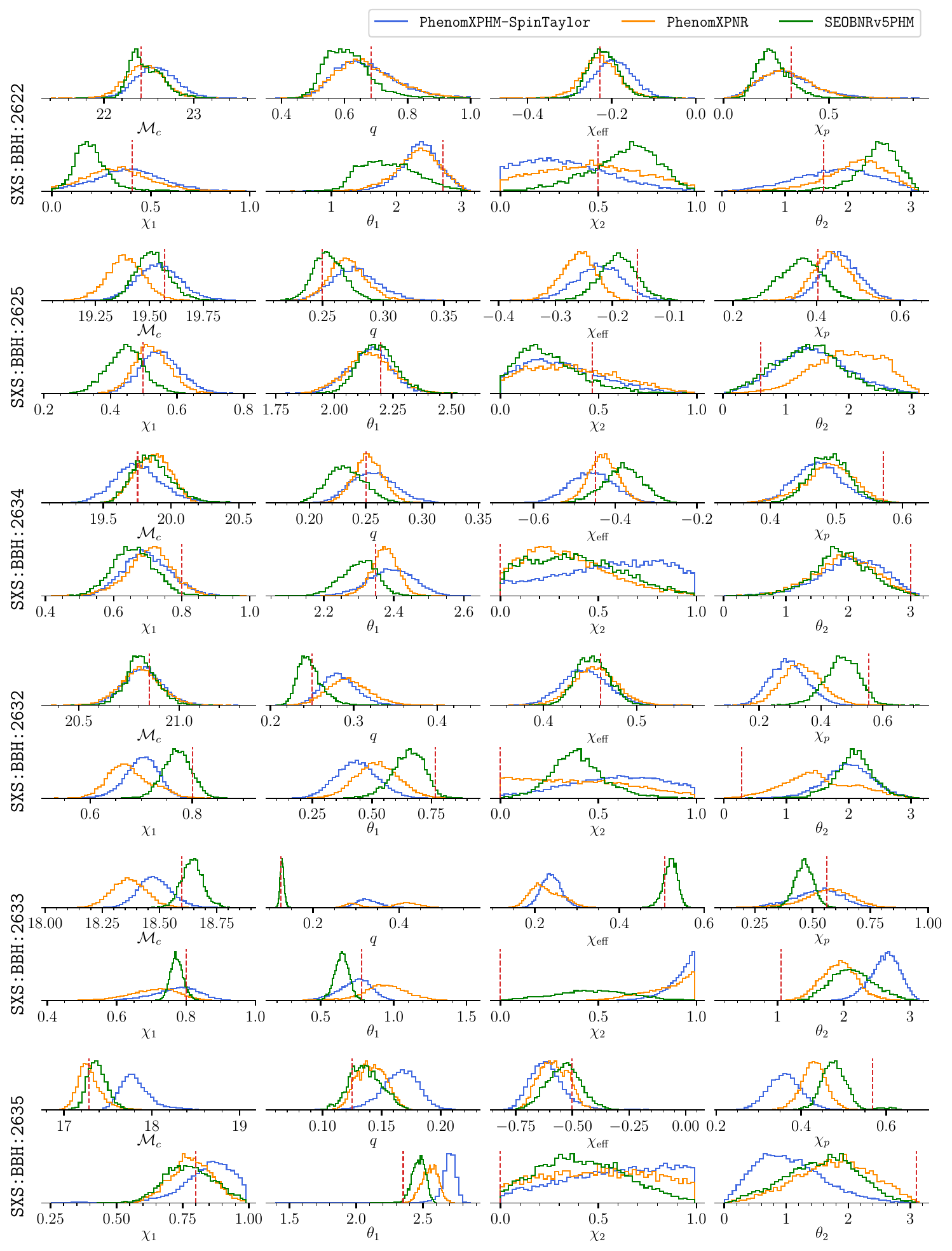}
  \caption{Systematic biases in GR recovery showing the inferred posteriors of the masses, spins, and spin angles for 6 long-duration SXS simulations of BBH systems. Each pair of consecutive rows corresponds to a single injection recovered with three different GR waveform models: \phXPHMST{} (blue), \phXPNR{} (orange), and \SEOBvFive{} (green). For each injection, the first row shows the chirp mass $\mathcal{M}_c$, mass ratio $q$, and effective-spin parameters parallel and perpendicular to the orbital angular momentum, $\chi_{\rm eff}$ and $\chi_p$, respectively. The second row shows the component spin magnitudes $\chi_i$ and their tilt angles $\theta_i$ with respect to the orbital angular momentum. The injected values are indicated by the red dashed lines. The remaining six simulations of Table~\ref{tab:sxs_simulations} are
shown in Fig.~\ref{fig:NRInjsGR2} (Appendix~\ref{app:GR}).}
  \label{fig:NRInjsGR}
\end{figure*}

\subsection{Biases in GR recovery}

We first perform the parameter estimation analyses under the assumption of GR using three state-of-the-art precessing waveform models: \phXPHMST, \phXPNR, and \SEOBvFive. Fig.~\ref{fig:NRInjsGR} shows the inferred 1D marginalized posteriors of the source parameters. For each injection, the posteriors are shown in two consecutive rows. The first row displays the chirp mass $\mathcal{M}_c$, mass ratio $q$, and effective-spin parameters parallel and perpendicular to the orbital angular momentum, $\chi_{\rm eff}$ and $\chi_p$, respectively. The second row shows the component spin magnitudes $a_i$ and their tilt angles $\theta_i$ with respect to the orbital angular momentum.

Since the injections are performed in zero noise, any offset between the posterior and the injected value reflects a bias. Small offsets of the marginalized posteriors can also arise from degeneracies with other binary parameters and from prior boundaries, whereas larger offsets, where the injected value lies in the tail of or outside the posterior, indicate waveform systematics.

\paragraph{Mass parameters:} The chirp mass shows only small offsets for most injections, with the injected value lying within the bulk of the posterior. Larger biases appear for \sxsid{2625}, for which \phXPNR{} underestimates $\mathcal{M}_c$, and for the two $Q=8$ systems discussed below. The mass ratio shows larger shifts. For \sxsid{2622}, all three models underestimate $q$, most strongly \SEOBvFive{}. For the $Q=4$ systems, the shifts are moderate and model dependent: the Phenom models favour larger $q$ for \sxsid{2625} and \sxsid{2632}, whereas \SEOBvFive{} favours a smaller $q$ for \sxsid{2634}. The largest biases are found for the most asymmetric systems. For \sxsid{2633}, \phXPHMST{} and \phXPNR{} favour a much less asymmetric binary and underestimate $\mathcal{M}_c$, whereas \SEOBvFive{} shows only a small offset. For \sxsid{2635}, \phXPHMST{} shifts both $\mathcal{M}_c$ and $q$ towards larger values, while the offsets for \phXPNR{} and \SEOBvFive{} are small. Mass-ratio biases therefore become more important for highly asymmetric systems, although no model performs uniformly better for all configurations.

\paragraph{Individual spin magnitudes and tilt angles:} The biases are more pronounced in the individual spins. For \sxsid{2622}, \SEOBvFive{} favours a smaller primary-spin magnitude $\chi_1$ and a substantially smaller tilt angle $\theta_1$, while the two Phenom models show smaller offsets. For \sxsid{2625}, the offsets in the primary spin are small, but the secondary-spin magnitude $\chi_2$ is underestimated. For \sxsid{2634}, all three models underestimate $\chi_1$, and for \sxsid{2632}, all three underestimate $\theta_1$, most strongly the Phenom models, which also favour a smaller $\chi_1$, i.e.\ a more aligned configuration. In both cases, the models allow nonzero values of $\chi_2$, even though the secondary BH is nonspinning in the injections. The strongest model dependence is found for \sxsid{2633}, where the Phenom models favour a large secondary spin, while \SEOBvFive{}, despite small offsets in the masses and $\chi_{\rm eff}$, underestimates $\theta_1$. For \sxsid{2635}, all three models overestimate $\theta_1$, most strongly \phXPHMST{}, which also pushes $\chi_1$ towards larger values. For injections with $\chi_2 = 0$, the corresponding angle $\theta_2$ is not physically defined, and its posterior should therefore not be interpreted as a bias.

\begin{figure*}[htbp]
\centering
\includegraphics[width=0.97\textwidth]{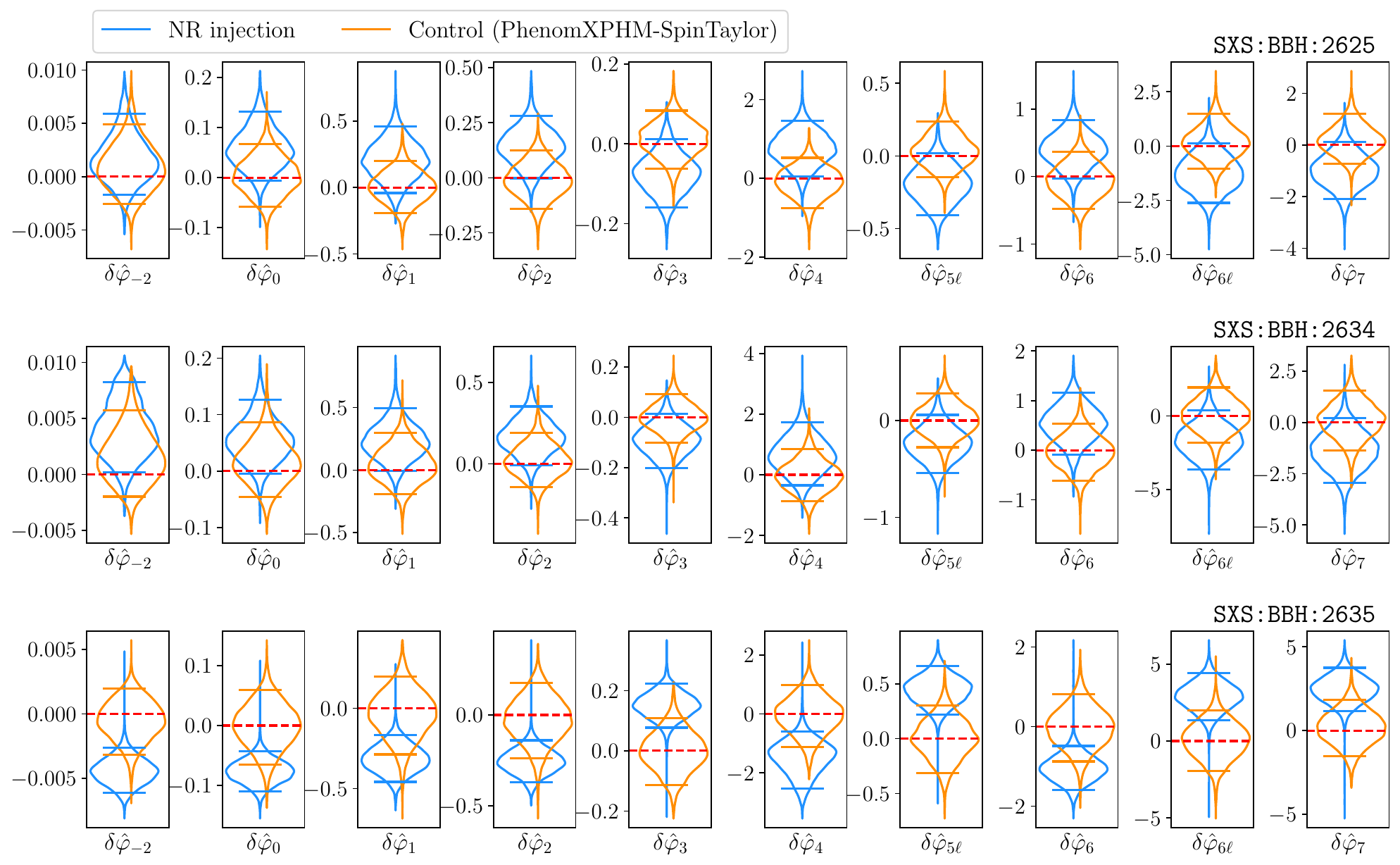}
  \caption{Systematic biases in the TIGER inspiral test from NR injections, showing the marginalized posterior distributions of the parameterized inspiral deviation coefficients for \sxsid{2625}, \sxsid{2634}, and \sxsid{2635}, whose parameters are listed in Table~\ref{tab:sxs_simulations}. Blue distributions correspond to NR injections recovered with \phXPHMST, while orange distributions show control injections generated and recovered with \phXPHMST. The latter provide a reference for biases in the absence of waveform-model disagreement. The horizontal dashed lines denote the GR prediction, $\delta\hat{\varphi}_i=0$, and horizontal bars indicate the corresponding 90\% credible intervals. }
  \label{fig:NRInjsTIGER}
\end{figure*}
\begin{figure*}[htbp]
\centering
  \includegraphics[width=0.97\textwidth]{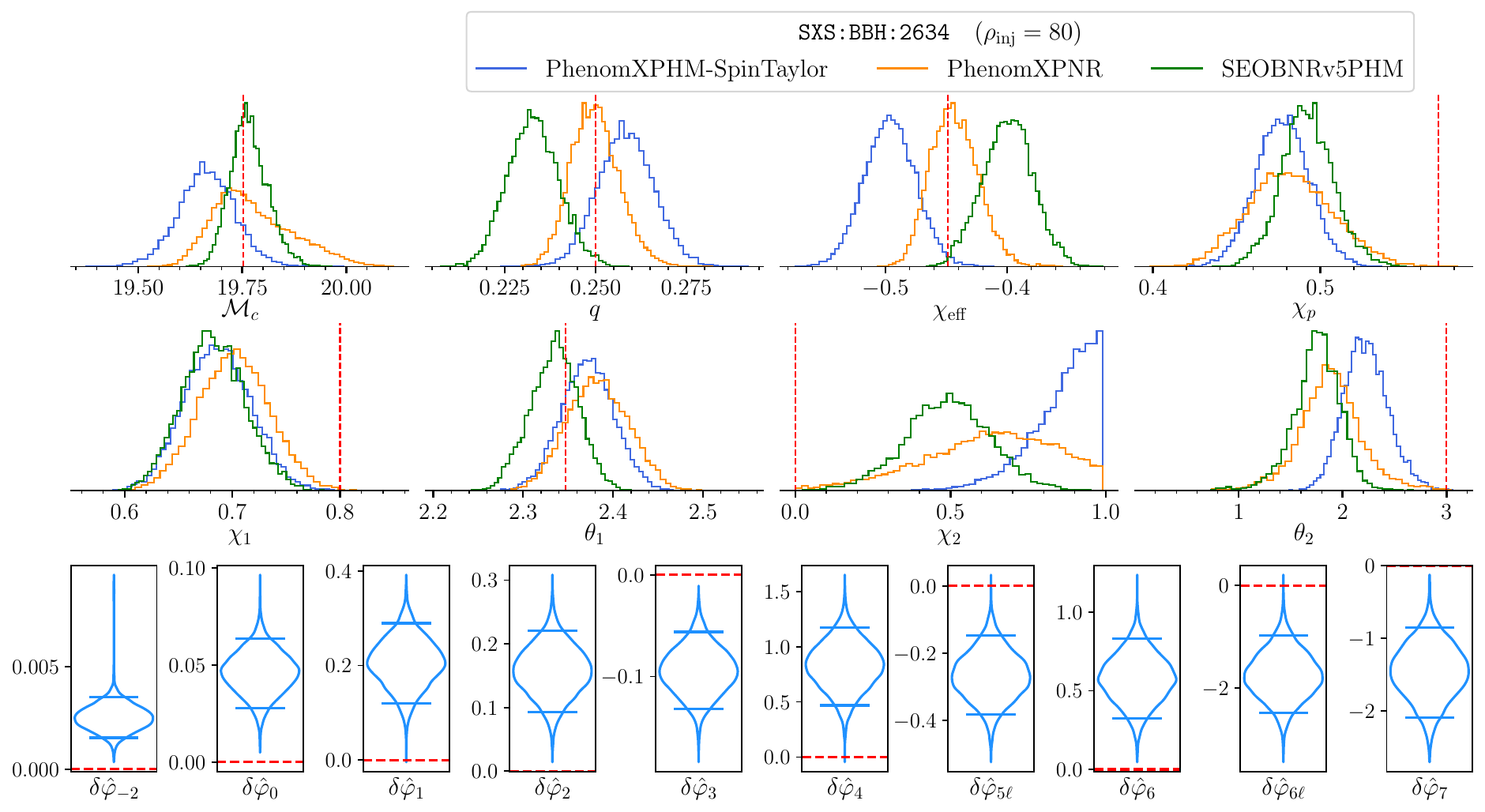}
  \caption{Systematic biases in the GR and TIGER analyses for a loud injection of the NR simulation \sxsid{2634} with optimal SNR $\rho_{\rm inj} = 80$; all other injection parameters are the same as in Fig.~\ref{fig:NRInjsTIGER}. The top and middle rows show the marginalized posteriors of the binary parameters obtained with three different GR waveform models: \phXPHMST (blue), \phXPNR (orange), and \SEOBvFive (green). The bottom row shows the marginalized posteriors of the PN deviation parameters obtained from TIGER analyses with \phXPHMST, where each coefficient is varied individually. The red dashed lines indicate the injected values (top and middle rows) and the GR prediction, $\delta\hat{\varphi}_i = 0$ (bottom row), and the horizontal bars denote the 90\% credible intervals.}
  \label{fig:NRInjsTIGERloud}
\end{figure*}

\paragraph{Effective spins:} The effective-spin parameters provide a useful summary of these individual-spin shifts. The aligned effective spin $\chi_{\rm eff}$ shows only small offsets for all three models for \sxsid{2622} and \sxsid{2632}, and for the Phenom models for \sxsid{2634}, even when noticeable biases are present in $\chi_i$ and $\theta_i$. For \sxsid{2632}, for example, the smaller $\chi_1$ and $\theta_1$ favoured by the Phenom models largely compensate in $\chi_{\rm eff}$ while reducing $\chi_p$. Correlated changes in the spin magnitudes and orientations can therefore preserve the projected-spin combination while giving an inaccurate reconstruction of the individual spins. For the other systems, $\chi_{\rm eff}$ is biased: all three models favour more negative values for \sxsid{2625} and \sxsid{2635}, \SEOBvFive{} favours a less negative value for \sxsid{2634}, and for \sxsid{2633} the two Phenom models strongly underestimate $\chi_{\rm eff}$, while \SEOBvFive{} shows only a small offset. The precession parameter $\chi_p$ is underestimated by all three models for \sxsid{2632}, \sxsid{2634}, and \sxsid{2635}, and \SEOBvFive{} underestimates it for all six injections. A small bias in $\chi_{\rm eff}$ therefore does not necessarily imply an accurate recovery of the precessing-spin dynamics.


\subsection{Biases in inspiral parameterized test}
For the inspiral parameterized test, we perform TIGER analyses using the NR simulations as injections and \phXPHMST{} as the recovery waveform model. Fig.~\ref{fig:NRInjsTIGER} shows the marginalized posteriors of the inspiral deviation parameters for three representative NR injections, selected to investigate the effects of spin magnitude and mass ratio. \sxsid{2625} and \sxsid{2634} share the same mass ratio, $Q=4$, but differ in spin magnitude, while \sxsid{2634} and \sxsid{2635} have nearly identical spin configurations and differ mainly in mass ratio. To distinguish waveform-systematic effects from biases intrinsic to the parameterized inference, we also perform a corresponding set of GR injections generated with \phXPHMST{} and recover them with the same waveform model. These control analyses are shown in orange, while the NR injections are shown in blue. Since the injection and recovery models are identical for the control cases, their posteriors provide a reference for the level of bias expected in the absence of waveform-model disagreement.


The control injections are consistent with GR, with posteriors peaking close to $\delta\hat{\varphi}_i = 0$. In contrast, the NR injections show clear, configuration-dependent shifts away from zero. For \sxsid{2625} and \sxsid{2634}, both with $Q=4$, most coefficients are displaced, but the GR value generally remains within or close to the edge of the 90\% credible interval, and the deviations therefore remain marginal at this SNR. Interestingly, the two systems are affected at different PN orders: for \sxsid{2625}, the GR value lies well within the credible interval for the lowest-order coefficients but moves to the edge from 1PN onwards, whereas for \sxsid{2634} the GR value lies at the edge for the low-order coefficients and well inside for the higher-order ones. Notably, the shifts for \sxsid{2625}, with moderate spin magnitudes $|\vec{\chi}_i| \simeq 0.5$, are comparable to those for \sxsid{2634}, whose primary spin is $0.8$, showing that the size of the bias is not set by the spin magnitude alone.

The most asymmetric system, \sxsid{2635} with $Q=8$, shows the largest effect: for all ten inspiral coefficients, the GR value lies outside the 90\% credible interval of the NR injection, while the corresponding control posteriors remain consistent with zero. This is also the system for which the GR recovery with \phXPHMST shows the largest shifts in $\mathcal{M}_c$ and $q$ (Fig.~\ref{fig:NRInjsGR}), suggesting that the bias in the source parameters propagates into the parameterized test.

Within each injection, the shifts follow a coherent sign pattern across PN orders, which is reversed for \sxsid{2635} compared with the two $Q=4$ systems. Since each coefficient is varied individually, this pattern reflects the same modelling residual being absorbed by different deviation parameters through their correlations with each other and with the chirp mass, rather than independent deviations at several PN orders.

\subsection{Loud signals}
\label{subsec:loud}
Waveform systematics become more important as the SNR increases, since the statistical uncertainty decreases while the modelling error remains. To illustrate this, we repeat the analysis of \sxsid{2634} with an optimal SNR of 80, keeping all other injection parameters fixed. Fig.~\ref{fig:NRInjsTIGERloud} shows the GR recoveries with the three waveform models (top and middle rows) and the TIGER posteriors obtained with \phXPHMST (bottom row).

In the GR analyses, the three models now give clearly separated posteriors. \phXPHMST underestimates $\mathcal{M}_c$ and overestimates $q$, while \SEOBvFive favours a smaller $q$. For $\chi_{\rm eff}$, \phXPHMST and \SEOBvFive lie on opposite sides of the injected value, with \phXPNR closest to it. All three models underestimate $\chi_p$ and $\chi_1$, and none recovers the nonspinning secondary; \phXPHMST even pushes $\chi_2$ to the upper prior boundary. These offsets are now large compared with the statistical width of the posteriors.

In the TIGER analysis, the GR value lies outside the 90\% credible interval for all ten inspiral coefficients. The shifts follow the same sign pattern as they did at an SNR of 30 (Fig.~\ref{fig:NRInjsTIGER}), and their central values are similar, while the credible intervals are narrower by roughly a factor of three to four. The same system that was only marginally inconsistent with GR at an SNR of 30 would therefore be interpreted as a strong violation at an SNR of 80. This shows that for loud events, waveform systematics alone can produce an apparently significant deviation from GR.

\section{Conclusion}
\label{sec:conclusion}
In this work, we studied the promise and pitfalls of highly spinning BBHs for parameterized tests of GR. Using controlled injections, we found that spin orientation strongly affects the measurability of the deviation parameters. For aligned spins, the 2PN parameter $\dphi_4$ remains strongly correlated with the primary-spin magnitude, even for $q=1/6$, whereas spin precession substantially reduces this degeneracy, with the tightest constraints for $\theta_1=\pi/3$. At $-1$PN and 3PN, positive projected spins tighten the constraints relative to negative projected spins, while the 2PN coefficient shows no monotonic trend. For the anti-aligned configurations with $q=1/2$ and $q=1/4$, the apparent deviations are substantially reduced by adopting a uniform-tilt prior. For $q=1/6$, a degeneracy between $\delta\hat\varphi_4$ and the intrinsic parameters remains under the same prior.

Using long SXS simulations, we then showed that current precessing waveform models produce model-dependent biases in the recovered source parameters. The largest mass biases appear for the most asymmetric systems, the individual spins are more strongly biased than $\chi_{\rm eff}$, and $\chi_p$ is frequently underestimated. These biases are not restricted to extreme configurations: even for the nearly comparable-mass, moderately spinning system \sxsid{2622}, all three models underestimate the mass ratio, and \SEOBvFive{} substantially underestimates the primary-spin magnitude, tilt angle, and $\chi_p$, while $\mathcal{M}_c$ and $\chi_{\rm eff}$ show only small offsets.

In the TIGER analyses, these waveform inaccuracies shift the deviation parameters away from zero, while the corresponding control injections remain consistent with GR. At SNR of 30, the shifts are marginal for the $Q=4$ systems, but GR lies outside the 90\% credible interval for all inspiral coefficients for the $Q=8$ system. The bias is not set by the spin magnitude alone, since the moderately spinning \sxsid{2625} shows shifts comparable to those of \sxsid{2634}. For a loud injection with SNR of 80, the same shifts remain while the credible intervals shrink, turning a marginal inconsistency into a strong false violation of GR.

Together, these results show that the promise and the pitfalls of highly spinning binaries are closely connected: the asymmetric, precessing configurations that provide the most sensitive tests are also those for which current waveform models are least accurate.

Our results also suggest practical diagnostics for separating a genuine GR violation from waveform systematics. First, a physical deviation should be consistent across independent waveform families, whereas the systematic shifts found here are strongly model dependent (Figs.~\ref{fig:NRInjsGR} and~\ref{fig:NRInjsTIGERloud}), although agreement between models is not sufficient on its own, since approximants can share common assumptions. Second, the expected systematic bias for a given event can be estimated with NR or NR surrogate~\cite{Varma:2019csw} injections near the inferred source parameters, analyzed with the same setup as the data, in the same way as the comparison between NR and control injections in Fig.~\ref{fig:NRInjsTIGER}. Third, for nearly anti-aligned configurations, the robustness of the result should be checked against the choice of spin-tilt prior (Fig.~\ref{fig:antialigned_prior}). At the catalog level, a correlation between the inferred deviations and mass ratio or precession would point to waveform systematics rather than new physics. The pattern of deviations across PN orders alone, however, is not a reliable diagnostic, since a deviation at a single order also propagates into the other coefficients through their correlations with each other and with the chirp mass.

Beyond tests of GR, the biases shown in Fig.~\ref{fig:NRInjsGR} can also affect population studies, where the component masses, spin magnitudes, and spin--orbit tilt angles of BBH populations~\cite{LIGOScientific:2025pvj} are used to distinguish isolated binary evolution from dynamical formation~\cite{Rodriguez:2016vmx} and to identify possible hierarchical mergers~\cite{Gerosa:2021mno}. Biases in $q$, $\chi_i$, $\theta_i$, $\chi_{\rm eff}$, or $\chi_p$ can therefore alter the inferred contributions of different formation channels and strengthen or weaken the evidence for second-generation black holes~\cite{Dhani:2024jja,Hoy:2026izf}. In particular, an underestimated $\chi_p$ may make a genuinely precessing binary appear less precessing, while a spurious secondary spin may lead to an incorrect interpretation of the spin configuration. Since waveform-induced biases can be coherent across events, their impact on population inference grows with catalog size, as recently shown for eccentric waveform models~\cite{Zeeshan:2026vsi}, and is expected to become more severe for next-generation detectors~\cite{Kapil:2024zdn}. Waveform inaccuracies can similarly propagate into standard-siren measurements of the Hubble constant through correlations with the luminosity distance and binary inclination~\cite{Dhani:2025xgt}.

Our analysis is based on zero-noise injections, so the shifts reported here represent the systematic offset expected on average; in real noise, the posteriors would additionally scatter around these offsets. Although we focused on TIGER, other tests of GR used in recent LVK analyses~\cite{LIGOScientific:2026oim} also rely on GR waveform models and are expected to be affected as well. The FTI test~\cite{Mehta:2022pcn} adopts a non-precessing baseline model, so that neglected precession would add to the systematics found here~\cite{Chandramouli:2024vhw}. Tests that directly probe spin-dependent effects, such as the spin-induced quadrupole moment test~\cite{Krishnendu:2017shb}, are particularly sensitive to the spin biases found in Sec.~\ref{sec:NRInjections}. These waveform-systematic effects are also expected to propagate to principal-component-based parameterized tests~\cite{Saleem:2021nsb, Datta:2022izc, Mahapatra:2025cwk}, since the principal components are constructed from combinations of the same underlying deviation parameters and can therefore inherit biases present in them. The inspiral--merger--ringdown consistency test~\cite{Ghosh:2016qgn, Ghosh:2017gfp} has already been shown to produce apparent deviations due to waveform systematics~\cite{Akcay:2025rve}, in particular from mismodelling the ringdown of precessing binaries with high mass ratios and large, strongly tilted spins~\cite{Foo:2024exr}, and has recently been excluded from LVK analyses because of such spurious biases~\cite{LIGOScientific:2026oim}. Merger--ringdown tests such as pSEOBNR~\cite{Maggio:2022hre, Pompili:2025cdc, Grimaldi:2026prn} are built on \SEOBvFive, for which we also find model-dependent biases in Fig.~\ref{fig:NRInjsGR}.  Similarly, tests of propagation effects based on a modified dispersion relation~\cite{Baka:2025drk} introduce phase corrections with the same frequency dependence as the PN terms, so that biases in the source parameters may also propagate into these tests. The same issue extends to searches with parameterized non-GR templates~\cite{Narola:2022aob, Sharma:2023djw}, where a mismodelled GR signal from a highly spinning, precessing binary could be recovered with a higher SNR by non-GR templates, mimicking an exotic signal. Quantifying these effects for highly spinning, precessing binaries will be important before interpreting any apparent deviation from GR in current and future observing runs.



\section{acknowledgments}
This work was supported by the Fonds de la Recherche Scientifique (F.R.S.-FNRS) under the funding IISN-4.4501.19. J.J. also acknowledges support from the Royal Observatory of Belgium. M.H. is supported by the Institutes Organisation of the Dutch Research Council (NWO-I) through Nikhef. We have used \numpy~\cite{Harris:2020xlr}, \scipy~\cite{Virtanen:2019joe}, \matplotlib~\cite{Hunter:2007ouj} for analyses and preparing the figures in the manuscript. Part of this work was carried out using computational resources provided by the LIGO Laboratory and supported by the National Science Foundation Grants PHY-0757058 and PHY-0823459. The authors are grateful for this support. This research has made use of data or software obtained from the Gravitational Wave Open Science Center (gwosc.org), a service of the LIGO Scientific Collaboration, the Virgo Collaboration, and KAGRA. This material is based upon work supported by NSF's LIGO Laboratory which is a major facility fully funded by the National Science Foundation, as well as the Science and Technology Facilities Council (STFC) of the United Kingdom, the Max-Planck-Society (MPS), and the State of Niedersachsen/Germany for support of the construction of Advanced LIGO and construction and operation of the GEO600 detector. Additional support for Advanced LIGO was provided by the Australian Research Council. Virgo is funded, through the European Gravitational Observatory (EGO), by the French Centre National de Recherche Scientifique (CNRS), the Italian Istituto Nazionale di Fisica Nucleare (INFN) and the Dutch Nikhef, with contributions by institutions from Belgium, Germany, Greece, Hungary, Ireland, Japan, Monaco, Poland, Portugal, Spain. KAGRA is supported by Ministry of Education, Culture, Sports, Science and Technology (MEXT), Japan Society for the Promotion of Science (JSPS) in Japan; National Research Foundation (NRF) and Ministry of Science and ICT (MSIT) in Korea; Academia Sinica (AS) and National Science and Technology Council (NSTC) in Taiwan.

\appendix

\section{Biases in GR recovery of NR injections }
\label{app:GR}
\begin{figure*}[htbp]
\centering
  \includegraphics[width=0.97\textwidth]{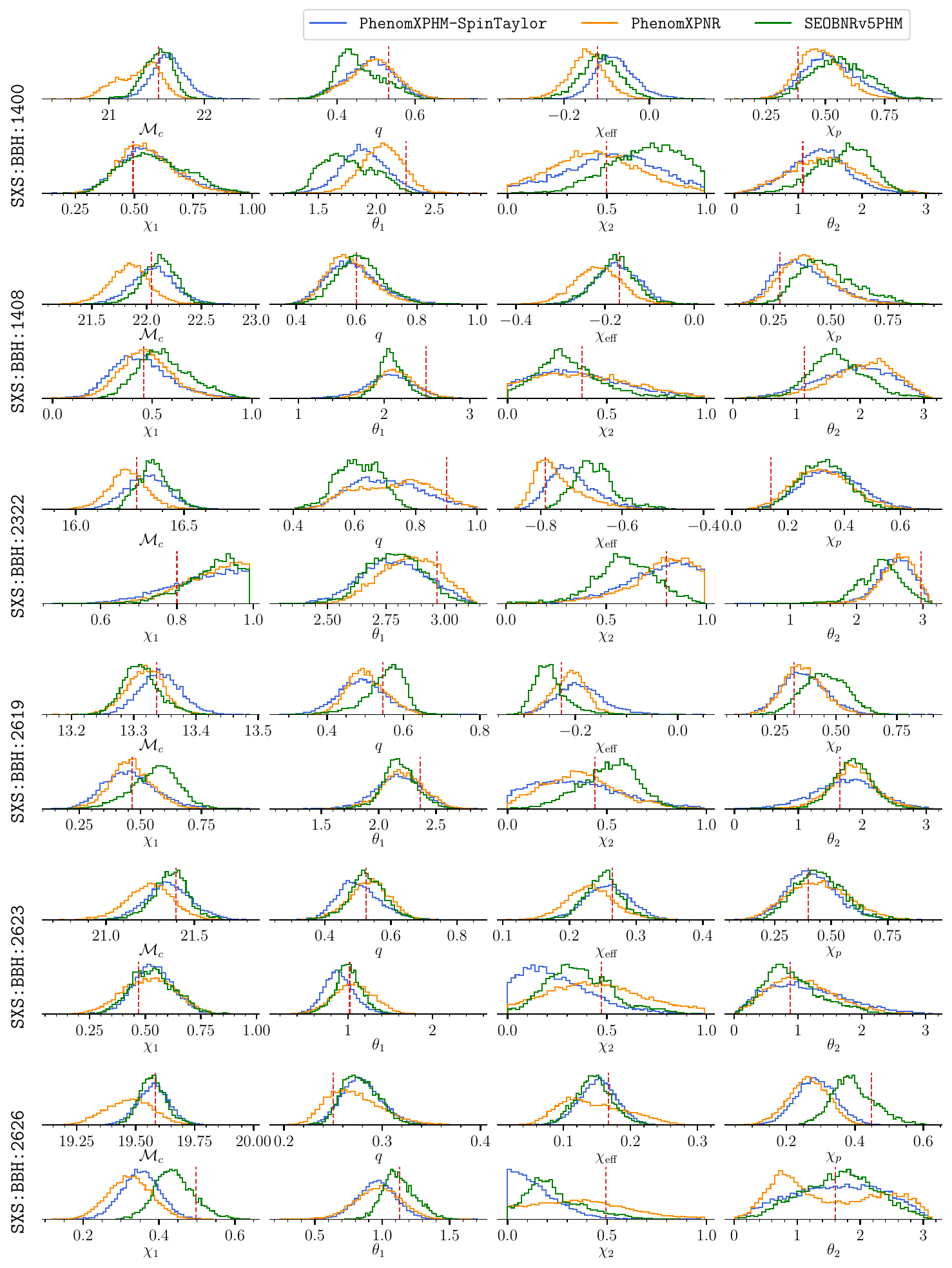}
  \caption{Same as Fig.~\ref{fig:NRInjsGR}, for the remaining six SXS simulations
listed in Table~\ref{tab:sxs_simulations}.}
  \label{fig:NRInjsGR2}
\end{figure*}

Figure~\ref{fig:NRInjsGR2} shows the GR recovery for the remaining six SXS simulations listed in Table~\ref{tab:sxs_simulations}, using the same setup as in Sec.~\ref{sec:NRInjections}. Four of them, \sxsid{1400}, \sxsid{1408}, \sxsid{2619}, and \sxsid{2623}, are comparable-mass systems ($Q \simeq 1.7$--$1.9$) with moderate spins. For these systems, $\mathcal{M}_c$, $q$, and $\chi_{\rm eff}$ show only small to moderate, model-dependent offsets, such as the smaller $q$ favoured by \SEOBvFive{} for \sxsid{1400}. Larger offsets appear in the spin orientations: all three models underestimate $\theta_1$ for \sxsid{1400} and \sxsid{1408}, and mildly for \sxsid{2619}, and overestimate $\chi_p$ for \sxsid{1400} and \sxsid{1408}, partly reflecting the weak constraints on $\chi_p$ for comparable-mass binaries. \sxsid{2623}, the only one of these systems with positive $\chi_{\rm eff}$, is the best recovered, with the injected values lying within the bulk of the posteriors. The longest simulation, \sxsid{2619} with 143 orbits, does not show larger biases than the shorter ones.

\sxsid{2322} is a nearly equal-mass binary with both spins close to anti-alignment ($\chi_{\rm eff} = -0.79$). All three models pull the tilt angles away from the anti-aligned configuration, overestimate $\chi_p$, and push $\chi_1$ towards the upper prior boundary, consistent with the prior-volume effect near $\theta_1 = \pi$ discussed in Sec.~\ref{subsec:antialigned}. In addition, all three models underestimate $q$, most strongly \SEOBvFive{}, which also underestimates $\chi_2$ and favours a less negative $\chi_{\rm eff}$.

Finally, \sxsid{2626} has the same mass ratio, $Q=4$, and similar spin magnitudes as \sxsid{2625}, but positive $\chi_{\rm eff}$. As for the $Q=4$ systems in Fig.~\ref{fig:NRInjsGR}, $\chi_p$ is underestimated, here most strongly by the two Phenom models, which also underestimate $\chi_1$, while $\mathcal{M}_c$ and $\chi_{\rm eff}$ show only small offsets. Overall, these results are consistent with Sec.~\ref{sec:NRInjections}: the masses and $\chi_{\rm eff}$ are generally recovered more accurately than the individual spins and their orientations, and the size and direction of the biases depend on both the waveform model and the binary configuration.

\clearpage

\bibliographystyle{apsrev4-2-titles}
\bibliography{reference}

\end{document}